\documentclass{llncs}
\usepackage[T1]{fontenc}
\usepackage[utf8x]{inputenx}
\usepackage[british]{babel}

\usepackage{graphicx}
\usepackage{xspace}
\usepackage{amsmath}
\usepackage{amssymb}
\usepackage{color}
\usepackage{multirow}
\usepackage{url}
\usepackage{hyperref}
\usepackage{cleveref}
\usepackage{tikz}
\usepackage{scalefnt}
\usepackage{array}
\usepackage{enumitem}
\usepackage{upgreek}
\usepackage{arydshln}

\setitemize{noitemsep,topsep=0pt,parsep=0pt,partopsep=0pt}

\usepackage{tikz}
\usetikzlibrary{calc,automata,positioning,decorations.pathreplacing}
\tikzstyle{every picture}=[shorten >=0pt]
\tikzstyle{every initial by arrow}=[initial text=]
\tikzstyle{every state}=[minimum size=14pt]
\tikzstyle{every node}=[font=\footnotesize,thick]
\tikzset{align at top/.style={baseline=(current bounding box.north)}}
\usepackage[outline]{contour}
\contourlength{1.5pt}

\usepackage[vlined,linesnumbered]{algorithm2e}
\SetAlCapFnt{\small}
\SetAlCapSkip{2ex plus 0.5ex minus 1ex}
\SetEndCharOfAlgoLine{}
\SetArgSty{textrm}
\SetKwInOut{Input}{Input}\SetKwInOut{Output}{Output}
\SetCommentSty{textit}
\SetKw{Break}{break}
\SetKwProg{With}{with}{:}{end}
\SetKwProg{fn}{function}{}{}
\SetKwFor{WhileTrue}{repeat}{}{end}
\SetKw{Continue}{continue}
\SetKwRepeat{DoWhile}{repeat}{while}
\crefname{algocf}{algorithm}{algorithms}
\Crefname{algocf}{Algorithm}{Algorithms}

\newcommand{\modest}{\textsc{\mbox{Modest}}\xspace}
\newcommand{\mcsta}{\textsf{\mbox{mcsta}}\xspace}

\newcommand{\toolset}{\textsc{\mbox{Modest} Toolset}\xspace}
\newcommand{\eg}{e.g.\ }
\newcommand{\ie}{i.e.\ }
\newcommand{\st}{\ifmmode \:\text{s.t.}\ \else s.t.\xspace\fi}
\newcommand{\etal}{et al.\xspace}
\newcommand{\wrt}{w.r.t.\xspace}
\newcommand{\sunit}[1]{\text{\begin{scriptsize}\,#1\end{scriptsize}}}
\newcommand{\ssunit}[1]{\text{\begin{scriptsize}#1\end{scriptsize}}}
\newcommand{\prism}{\textsc{Prism}\xspace}

\newcommand{\Dist}[1]{\ensuremath{\mathrm{Dist}({#1})}\xspace}
\newcommand{\support}[1]{\ensuremath{\mathrm{support}({#1})}\xspace}
\newcommand{\set}[1]{\ensuremath{\{\,#1\,\}}}

\newcommand{\tuple}[1]{\ensuremath{\langle #1 \rangle}}
\newcommand{\powerset}[1]{\ensuremath{2^{#1}}\xspace}
\newcommand{\RR}{\ensuremath{\mathbb{R}}\xspace}
\newcommand{\NN}{\ensuremath{\mathbb{N}}\xspace}
\newcommand{\True}[0]{\ensuremath{\mathit{true}}\xspace}
\newcommand{\False}[0]{\ensuremath{\mathit{false}}\xspace}

\newcommand{\sem}[1]{\ensuremath{[\![ #1 ]\!]}}

\begin{document}

\title{%
Explicit Model Checking of Very Large MDP \\using Partitioning and Secondary Storage%
\thanks{This work is supported by the EU 7th Framework Programme under grant agreements 295261 (MEALS) and 318490 (SENSATION), by the DFG as part of  SFB/TR 14 AVACS, by the CAS/SAFEA International Partnership Program for Creative Research Teams, and by the CDZ project CAP (GZ 1023).}%
}
\author{
Arnd Hartmanns \and
Holger Hermanns
}

\institute{
Saarland University -- Computer Science, Saarbr\"ucken, Germany}
\date{\today}
\maketitle

\begin{abstract}
The applicability of model checking is hindered by the state space explosion problem in combination with limited amounts of main memory.
To extend its reach, the large available capacities of secondary storage such as hard disks can be exploited.
Due to the specific performance characteristics of secondary storage technologies, specialised algorithms are required.
In this paper, we present a technique to use secondary storage for probabilistic model checking of Markov decision processes.
It combines state space exploration based on partitioning with a block-iterative variant of value iteration over the same partitions for the analysis of probabilistic reachability and expected-reward properties.
A sparse matrix-like representation is used to store partitions on secondary storage in a compact format.
All file accesses are sequential, and compression can be used without affecting runtime.
The technique has been implemented within the \toolset.
We evaluate its performance on several benchmark models of up to 3.5~billion states.
In the analysis of time-bounded properties on real-time models, our method neutralises the state space explosion induced by the time bound in its entirety.
\end{abstract}

\section{Introduction}
\label{sec:Introduction}

Model checking~\cite{CGP99} is a formal verification technique to ensure that a given model of the states and behaviours of a safety- or performance-critical system satisfies a set of requirements.
We are interested in models that consider \emph{nondeterminism} as well as quantitative aspects of systems in terms of \emph{time} and \emph{probabilities}.
Such models can be represented as Markov decision processes (MDP~\cite{Put94}) and verified with \emph{probabilistic model checking}.
However, the applicability of model checking is limited by the state space explosion problem:
The number of states of a model grows exponentially in the number of variables and parallel components, yet they have to be represented in limited computer memory in some form.
Probabilistic model checking is particularly affected due to its additional numerical complexity.
Several techniques are available to stretch its limits:
For example, symbolic probabilistic model checking~\cite{AKNPS00}, implemented in the \prism tool~\cite{KNP11}, uses variants of binary decision diagrams (BDD) to compactly represent the state spaces of well-structured models in memory at the cost of verification runtime.
Partial order~\cite{BDG06} 
and confluence reduction~\cite{TSP11} 
deliver smaller-but-equivalent state spaces and work particularly well for highly symmetric models.
When trading accuracy for tractability or efficiency is acceptable, abstraction and refinement techniques like CEGAR~\cite{HWZ08} 
can be applied.
The common theme is that these approaches aim at reducing the state space or its representation such that it fits, in its entirety, into the main memory of the machine used for model checking.
An alternative is to store this data on secondary storage such as hard disks or solid state drives and only load small parts of it into main memory when and as needed.
This is attractive due to the vast difference in size between main memory and secondary storage:
Typical workstations today possess in the order of 4-8\,GB of main memory, but easily 1\,TB or more of hard disk space.
Moreover, with the advent of dynamically scalable cloud storage, virtually unlimited off-site secondary storage has become easily accessible.
For conciseness, we from now on refer to main memory as \emph{memory} and to any kind of secondary storage as \emph{disk}.

In this paper, we present a method and tool implementation for disk-based probabilistic model checking of MDP.
Any such approach  must solve two tasks:
State space \emph{exploration}, the generation and storage on disk of a representation of the reachable part of the state space, and the disk-based \emph{analysis} to verify  the given properties of interest based on this representation.
The core challenge is that the most common type of secondary storage, magnetic hard disks, exhibits extremely low random-access performance, yet standard memory-based methods for exploration and analysis access the state space in a practically random way.

\subsubsection{Previous work.}
Exploration is an implicit graph search problem, and a number of solutions that reduce the amount of random accesses during search have been proposed in the literature.
These fall into three broad categories:
($i$)~exploiting the layered structure of breadth-first search (BFS) by keeping only the current BFS layer in memory while delaying duplicate detection \wrt previous layers until the current one has been fully explored~\cite{PITZ02,SD98};
($ii$)~partitioning the state space according to some given or automatically computed partitioning function over the states and then loading only one partition into memory at a time in an iterative process~\cite{BJ05,EK13};
($iii$)~treating memory purely as a cache for a disk-based search, but using clever hashing and hash partitioning techniques to reduce and sequentialise disk accesses~\cite{HW06}.
Exploration can naturally be combined on-the-fly with checking for the reachability of error states, and methods to perform on-the-fly verification of liveness and LTL properties exist~\cite{BBS07,EJ06,ESS08}.

The analysis of other logics, such as CTL model checking with satisfaction sets, and of other models, such as probabilistic model checking of MDP with value iteration, inherently require the entire state space for a dedicated analysis step following exploration.
Previous work on disk-based probabilistic model checking considers purely stochastic models and focusses on the analysis phase:
In absence of nondeterminism, classical block-iterative methods~\cite{Ste94} can be used  with disk-based (sparse) matrix representations of Markov models. They proceed by  loading into memory and analysing one matrix block at a time (plus those that it depends on) iteratively until the method has converged for all blocks.
Implementations can be divided into \emph{matrix-out-of-core} and \emph{complete out-of-core} approaches~\cite{Meh04}.
In the former, the vector of state values being iteratively computed is still kept in memory in its entirety~\cite{DS97}.
It is similar to how \prism~\cite{KNP11} uses BDD in its ``\textsc{hybrid}'' engine for the model only, while both model and values are represented symbolically in its ``\textsc{mtbdd}'' engine.
The symbolic and disk-based approaches for Markov chains can be combined~\cite{KMNP02}.
Further work on the disk-based analysis of purely stochastic models includes different implementations that are both disk-based and parallelised or distributed~\cite{BH06,HK99}.

For the nondeterministic-probabilistic model of MDP that we are concerned with, the default scalable analysis algorithm used in model checking is value iteration, an iterative fixpoint method that updates the values of each state based on a function over the values of its immediate successors until all changes remain below a given error.
We are aware of only one explicitly disk-based approach to value iteration, which associates the values to the transitions instead of the states and is based on sequentially traversing two files containing the transitions that have been externally sorted by source and target states in each iteration~\cite{EJB07}.
However, external sorting is a costly operation, leading to high runtime.

The correctness of value iteration depends neither on the order in which the updates are performed nor on how many updates a state receives in one iteration.
This can be exploited to improve its performance by taking the graph structure of the underlying model into account to perform more updates for ``relevant'' states in a ``good'' order.
One such technique is topological value iteration~\cite{DG07}, based on a division of the MDP into strongly connected components.
More generally, this means that value iteration can also be performed in a block-iterative manner.

\subsubsection{Our contribution.}
The technique for disk-based probabilistic model checking of MDP that
we present in this paper is a complete out-of-core method.  It
combines the state space partitioning approach from disk-based search
with a block-iterative variant of value iteration based on a very
compact sparse matrix-like representation of the partitions on disk.
In light of the disk space available, compactness seems at first sight
to be a non-issue, but in fact is a crucial aspect due to the low throughput of hard disks compared to main memory. Based on a given partitioning function, our approach proceeds by first exploring the partitions of the state space using an explicit state representation while directly streaming the sparse matrix-like representation to disk.
When exploration is completed, the stored partitions are analysed using a block-based variant of value iteration:
It iterates in an outer loop over the partitions on disk, for each of which value iterations are performed in an inner loop until convergence.
All read and write operations on the files we generate on disk are sequential.
We can thus easily add compression, which in our experiments reduces the amount of disk space needed by a factor of up to 10 without affecting overall runtime.

Our method has been implemented by extending the \mcsta tool~\cite{HHH14} of the \toolset~\cite{HH14}.
The implementation currently supports the computation of reachability probabilities and expected accumulated rewards.
To the best of our knowledge, \mcsta is at this point the only publicly available tool that provides disk-based verification of MDP.
We have evaluated the approach and its implementation on five case studies.
The largest model we consider has 3.5~billion states.
It can be explored and analysed in less than 8~hours using no more than 2\sunit{GB} of memory and 30\sunit{GB} disk space.
Our technique is particularly efficient for the analysis of time-bounded properties on real-time extensions of MDP.
In these cases, the overhead of using the disk is small and the enormous state space explosion caused by the time bounds can be neutralised in its entirety.

\section{Preliminaries}
\label{sec:Preliminaries}

The central formal model that we use are Markov decision processes:

\begin{definition}
A \emph{probability distribution} over a countable set~$\varOmega$ is a function $\mu \in \varOmega \to [0, 1]$ such that $\sum_{\omega \in \varOmega}{\mu(\omega)} = 1$.
Its \emph{support} is $\support{\mu} = \set{s \in S \mid \mu(s) > 0}$.
We denote by $\Dist{\varOmega}$ the set of all probability distributions over~$\varOmega$.
\end{definition}

\begin{definition}
A \emph{Markov decision process}~(MDP) is a triple $\tuple{S, T, s_0}$ consisting of a countable set of \emph{states} $S$, a \emph{transition function} $T \in S \to \powerset{\Dist{S \times R}}$ for a countable subset $R \subsetneq \RR$ with $T(s)$ countable for all $s \in S$, and an \emph{initial state} $s_0 \in S$.
A \emph{partitioning function} for an MDP is a function $f \in S \to \set{ 1, \dots, k }$ for some $k \in \NN$ with $f(s_0) = 1$.
\end{definition}
For $s \in S$, we call $\mu \in T(s)$ a \emph{transition} of~$s$, and a pair $b = \tuple{s', r} \in \support{\mu}$ a \emph{branch} of~$\mu$, with $s'$ being the \emph{target state} of~$b$ and $r$ being the associated \emph{reward} value.
MDP support both nondeterministic and probabilistic choices:
A state can have multiple outgoing transitions, each of which leads into a probability distribution over pairs $\tuple{s, r}$.
A partitioning function~$f \in S \to \set{1, \dots, n}$, $n \in \NN$, divides the states of an MDP into partitions $P_i = \set{ s \in S \mid f(s) = i }$.
The \emph{partition graph} is the directed graph $\tuple{P, U}$ with nodes $P = \set{ P_i \mid 1 \leq i \leq k }$ and edges $U = \set{ \tuple{P_i, P_j} \mid i \neq j \wedge \exists\,s \in P_i, \mu \in T(s), \tuple{s', r} \in \support{\mu}\colon s' \in P_j }$.
It is \emph{forward-acyclic} if there is no $\tuple{P_i, P_j} \in U$ with $j < i$.

We are interested in~the probability of reaching certain states in an MDP and in the expected reward  accumulated when doing so.
Since an MDP may contain nondeterministic choices, these values are only well-defined under a \emph{scheduler}, which provides a recipe to resolve the nondeterminism.
The verification questions are thus:
Given a set of states $F \subseteq S$,
($i$) what is the maximum/minimum probability of eventually reaching a state in~$F$ over all possible schedulers (\emph{reachability probability}), and
($ii$) what is the maximum/minimum expected accumulated reward once a state in~$F$ is reached for the first time over all possible schedulers (\emph{expected reward})?
These quantities can be formally defined using the usual cylinder set construction for the paths of the MDP~\cite{FKNP11}.

The computation of these quantities is typically done using \emph{value iteration}, as shown in \Cref{alg:ValueIteration} for maximum reachability probabilities.
For the minimum case, we replace maximisation by minimisation in line~\ref{alg:ValueIteration:IterationStep}.
To compute expected rewards, a precomputation step is needed to determine those states from which $F$ is reachable with probability one and zero, respectively.
This can be done with straightforward fixpoint algorithms over the graph structure of the MDP~\cite{FKNP11}.

\begin{algorithm}[tp]
$\mathit{values} := \set{ s \mapsto 1 \mid s \in F} \cup \set{ s \mapsto 0 \mid s \in S \setminus F }$\tcp*{the value vector}
  \Repeat{$\mathit{error} < \epsilon$}{
    $\mathit{error} := 0$\;
    \ForEach{$s \in S \setminus F$}{
      $v_\mathit{new} := \max \left\{ \sum_{\tuple{s',r} \in \support{\mu}}{\mu(s') \cdot \mathit{values}(s') \mid \mu \in T(s)} \right\}$\;\label{alg:ValueIteration:IterationStep}
      \lIf{$v_\mathit{new} > 0$}{$\mathit{error} := \max\{\, \mathit{error}, |v_\mathit{new} - \mathit{values}(s)| / \mathit{values}(s) \,\}$\label{alg:MDPExhaustiveMCValueIteration:ErrorLine}}
      $\mathit{values}(s) := v_\mathit{new}$\;
    }
  }
  \Return{$\mathit{values}(s_0)$}
\caption{\small Value iteration to compute max.\ reachability probabilities}
\label{alg:ValueIteration}
\end{algorithm}

Using MDP directly to build models of complex systems is cumbersome.
Instead, higher-level formalisms such as \prism's guarded command language are used.
They add  to MDP variables that take values from finite domains.
In an \emph{MDP with variables} (VMDP),
each transition is associated with a \emph{guard}, a Boolean expression that disables the transition when it is \False.
The probabilities and reward values of the branches are given as real-valued arithmetic expressions.
Every branch has an \emph{update} that assigns new values (given as expressions) to the variables of the process.
The semantics of a VMDP~$M$ is the MDP $\sem{M}$ whose states are pairs $\tuple{s, v}$ of a state~$s$ of $M$ and a valuation~$v$ for the variables.
Transitions out of $s$ that are disabled according to $v$ do not appear in $\sem{M}$, and the valuations of a branch's targets are computed by applying the update of the branch to the valuation of the transition's source state.
 A partitioning function~$f$ for a VMDP can be determined by an upper-bounded arithmetic expression~$e$ with values in~\NN: $f(\tuple{s, v}) = e(v)$ where $e(v)$ is the evaluation of $e$ in~$v$.
The reachability set $F$ can likewise be characterised by a Boolean expression.

\subsubsection{Real-time extensions of MDP}
To model and analyse real-time systems, MDP can be extended with real-valued clock variables and state invariant expressions as in timed automata (TA~\cite{AD94}), leading to the model of probabilistic timed automata (PTA~\cite{KNSS02}).
A number of techniques are available to model-check PTA~\cite{NPS13}, but only the digital clocks approach~\cite{KNPS06} allows the computation of both reachability probabilities and expected rewards:
Clocks are replaced by bounded integer variables, and self-loop transitions are added to increment them synchronously as long as the state invariant is satisfied.
This turns the (finite) PTA into a (finite) VMDP.
The conversion preserves reachability probabilities and expected reward values whenever all clock constraints in the PTA are closed and diagonal-free.
However, the size of the final MDP is exponential in the number of clock variables and the maximum constants that they are compared to.

For timed models, we are also interested in \emph{time-bounded reachability}:
Ranging over all possible schedulers, what is the maximum/minimum probability of eventually reaching a state in~$F$ within at most $t$ time units?
These probabilities can be computed by adding a new clock variable~$x$ to the PTA that is never reset and computing the reachability probability for the set $F' = \set{\tuple{s, v} \mid s \in F \wedge v(x) \leq t }$ in the resulting digital clocks MDP~\cite{NPS13}.

A further extension of PTA are stochastic timed automata (STA~\cite{BDHK06}).
They allow assignments of the form $x := \textsc{sample}(D)$ to sample from (continuous) probability distributions $D$, \eg exponential or normal distributions, in updates.
This allows for stochastic delays, such as the exponentially-distributed sojourn times of continuous-time Markov chains, in addition to the nondeterministic delays of (P)TA.
A first model checking technique for STA has recently been described~\cite{HHH14} and implemented within the \mcsta tool of the \toolset~\cite{HH14}.
It works by abstracting assignments that use continuous distributions into finite-support probabilistic choices plus continuous nondeterminism, turning the STA into a PTA that can be analysed with \eg the digital clocks technique.

\section{Disk-Based State Space Exploration with Partitioning}
\label{sec:PartitionedExploration}

In this section, we describe the partitioned state space exploration approach that we use in our disk-based analysis technique for MDP.
We assume that the MDP to be explored is given in some compact description that can be interpreted as a VMDP, and a partitioning function~$f$ is given as an expression over its variables.
Disk-based exploration using partitioning has been the subject of previous work~\cite{BJ05,EK13}, so we focus on the novel aspect of generating a sparse matrix-like representation of the MDP on-the-fly during explicit-state exploration with low memory usage and in a compact format in a single file on disk.

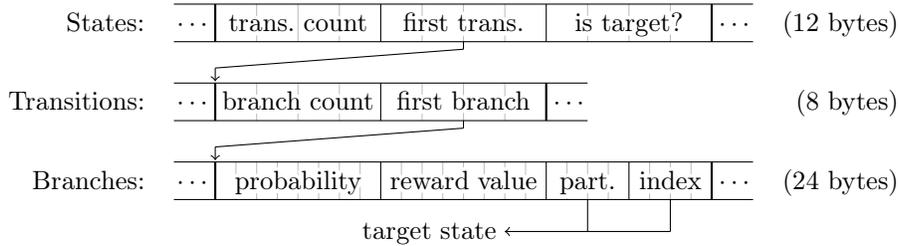
\begin{figure}[t]
\centering
\begin{tikzpicture}[text height=1.5ex,text depth=.25ex,x=1.10cm]
\tikzstyle{every node}=[font=\normalsize]
\draw (-0.25,-0.43) node [anchor=east] {States:};
\draw[thick]     (0.5,-0.15) -- (0.5,-0.65);
\draw[lightgray] (1.0,-0.15) -- (1.0,-0.65);
\draw[lightgray] (1.5,-0.15) -- (1.5,-0.65);
\draw[lightgray] (2.0,-0.15) -- (2.0,-0.65);
\draw[]          (2.5,-0.15) -- (2.5,-0.65);
\draw[lightgray] (3.0,-0.15) -- (3.0,-0.65);
\draw[lightgray] (3.5,-0.15) -- (3.5,-0.65);
\draw[lightgray] (4.0,-0.15) -- (4.0,-0.65);
\draw[]          (4.5,-0.15) -- (4.5,-0.65);
\draw[lightgray] (5.0,-0.15) -- (5.0,-0.65);
\draw[lightgray] (5.5,-0.15) -- (5.5,-0.65);
\draw[lightgray] (6.0,-0.15) -- (6.0,-0.65);
\draw[thick]     (6.5,-0.15) -- (6.5,-0.65);
\draw[] (0, -0.15) -- (7, -0.15);
\draw[] (0,-0.65) -- (7,-0.65);
\draw (0.25,-0.425) node [] {$\cdots$};
\draw (1.5,-0.43) node [] {\contour{white}{trans.\ count}};
\draw (3.5,-0.43) node [] {\contour{white}{first trans.}};
\draw (5.5,-0.43) node [] {\contour{white}{is target?}};
\draw (6.8,-0.425) node [] {$\cdots$};
\draw (7.25,-0.43) node [anchor=west] {(12 bytes)};
\draw (-0.25,-1.47) node [anchor=east] {Transitions:};
\draw[thick]     (0.5,-1.2) -- (0.5,-1.7);
\draw[lightgray] (1.0,-1.2) -- (1.0,-1.7);
\draw[lightgray] (1.5,-1.2) -- (1.5,-1.7);
\draw[lightgray] (2.0,-1.2) -- (2.0,-1.7);
\draw[]          (2.5,-1.2) -- (2.5,-1.7);
\draw[lightgray] (3.0,-1.2) -- (3.0,-1.7);
\draw[lightgray] (3.5,-1.2) -- (3.5,-1.7);
\draw[lightgray] (4.0,-1.2) -- (4.0,-1.7);
\draw[thick]     (4.5,-1.2) -- (4.5,-1.7);
\draw[] (0,-1.2) -- (5,-1.2);
\draw[] (0,-1.7) -- (5,-1.7);
\draw (0.25,-1.475) node [] {$\cdots$};
\draw (1.5,-1.47) node [] {\contour{white}{branch count}};
\draw (3.5,-1.47) node [] {\contour{white}{first branch}};
\draw (4.8,-1.475) node [] {$\cdots$};
\draw (7.25,-1.47) node [anchor=west] {\phantom{0}(8 bytes)};
\draw (-0.25,-2.53) node [anchor=east] {Branches:};
\draw[thick]     (0.50,-2.25) -- (0.50,-2.75);
\draw[lightgray] (0.75,-2.25) -- (0.75,-2.75);
\draw[lightgray] (1.00,-2.25) -- (1.00,-2.75);
\draw[lightgray] (1.25,-2.25) -- (1.25,-2.75);
\draw[lightgray] (1.50,-2.25) -- (1.50,-2.75);
\draw[lightgray] (1.75,-2.25) -- (1.75,-2.75);
\draw[lightgray] (2.00,-2.25) -- (2.00,-2.75);
\draw[lightgray] (2.25,-2.25) -- (2.25,-2.75);
\draw[] (2.50,-2.25) -- (2.50,-2.75);
\draw[lightgray] (2.75,-2.25) -- (2.75,-2.75);
\draw[lightgray] (3.00,-2.25) -- (3.00,-2.75);
\draw[lightgray] (3.25,-2.25) -- (3.25,-2.75);
\draw[lightgray] (3.50,-2.25) -- (3.50,-2.75);
\draw[lightgray] (3.75,-2.25) -- (3.75,-2.75);
\draw[lightgray] (4.00,-2.25) -- (4.00,-2.75);
\draw[lightgray] (4.25,-2.25) -- (4.25,-2.75);
\draw[]          (4.50,-2.25) -- (4.50,-2.75);
\draw[lightgray] (4.75,-2.25) -- (4.75,-2.75);
\draw[lightgray] (5.00,-2.25) -- (5.00,-2.75);
\draw[lightgray] (5.25,-2.25) -- (5.25,-2.75);
\draw[]          (5.50,-2.25) -- (5.50,-2.75);
\draw[lightgray] (5.75,-2.25) -- (5.75,-2.75);
\draw[lightgray] (6.00,-2.25) -- (6.00,-2.75);
\draw[lightgray] (6.25,-2.25) -- (6.25,-2.75);
\draw[thick]     (6.50,-2.25) -- (6.50,-2.75);
\draw[] (0,-2.25) -- (7.0,-2.25);
\draw[] (0,-2.75) -- (7.0,-2.75);
\draw (0.25,-2.525) node [] {$\cdots$};
\draw (1.5,-2.53) node [] {\contour{white}{probability}};
\draw (3.5,-2.53) node [] {\contour{white}{reward value}};
\draw (5.0,-2.53) node [] {\contour{white}{part.}};
\draw (6.0,-2.53) node [] {\contour{white}{index}};
\draw (6.8,-2.525) node [] {$\cdots$};
\draw (7.25,-2.53) node [anchor=west] {(24 bytes)};
\draw[] (3.50,-0.65) -- (3.50,-0.75);
\draw[] (3.50,-0.75) -- (0.50,-1.00);
\draw[->] (0.50,-1.00) -- (0.50,-1.175);
\draw[] (3.50,-1.70) -- (3.50,-1.80);
\draw[] (3.50,-1.80) -- (0.50,-2.05);
\draw[->] (0.50,-2.05) -- (0.50,-2.225);
\draw[] (5.00,-2.75) -- (5.00,-3.175);
\draw[] (6.00,-2.75) -- (6.00,-3.175);
\draw[->] (6.00,-3.175) -- (4.00,-3.175);
\draw[overlay] (4.0,-3.1875) node [anchor=east] {target state};
\end{tikzpicture}
\caption{In-memory representation of MDP for fast random access}
\label{fig:SparseMatrixMemory}
\end{figure}

\subsection{Representation of MDP in Memory and on Disk}

There are conceptually two ways to represent in memory an MDP that is the semantics of a VMDP:
In an \emph{explicit-state} manner, or in a \emph{sparse matrix-style} representation.
In the former, only the set of states of the MDP is kept, with each state stored as a vector $\tuple{s, v = \tuple{ v_1, \dots, v_n }}$ where $s$ identifies the state in the original VMDP and $v_{i}$ the value of its $i$-th variable.
Given a state and the compact description of the VMDP, we can recompute transitions and branches at any time on-demand.
The other alternative is to identify each of the $n$ states of the MDP with a value in $\set{1, \dots, n}$, its \emph{index}, and explicitly store the set of transitions belonging to a state index and the transitions' branches.
For each branch, its probability, its reward value, and the index of the target state need to be stored.
This sparse matrix-style representation takes its name from the similar idea of storing a Markov chain as a sparse encoding of its probability matrix.
All information about the inner structure of the states is discarded.

\Cref{fig:SparseMatrixMemory} outlines the sparse matrix-style representation used by \mcsta, which keeps three arrays to store the states, transitions and branches of a partition of the state space.
For a state, ``is target?'' is \True iff it is in the reachability set~$F$ that we consider.
The target state of a branch is identified by its partition and its relative index within that partition.
This format is more memory-efficient than an explicit-state representation when the model has many variables, and access to transitions and branches can be significantly faster because guards and other expressions in the model do not need to be evaluated on every access.

\begin{figure}[t]
\centering
\begin{tikzpicture}[text height=1.5ex,text depth=.25ex,x=1.175cm,decoration={brace,transform={yscale=1.5}}]
\tikzstyle{every node}=[font=\normalsize]
\draw[thick]     (0.25,0.25) -- (0.25,-0.25);
\draw[]          (0.50,0.25) -- (0.50,-0.25);
\draw[lightgray] (0.75,0.25) -- (0.75,-0.25);
\draw[lightgray] (1.00,0.25) -- (1.00,-0.25);
\draw[lightgray] (1.25,0.25) -- (1.25,-0.25);
\draw[lightgray] (1.50,0.25) -- (1.50,-0.25);
\draw[lightgray] (1.75,0.25) -- (1.75,-0.25);
\draw[lightgray] (2.00,0.25) -- (2.00,-0.25);
\draw[lightgray] (2.25,0.25) -- (2.25,-0.25);
\draw[]          (2.50,0.25) -- (2.50,-0.25);
\draw[lightgray] (2.75,0.25) -- (2.75,-0.25);
\draw[lightgray] (3.00,0.25) -- (3.00,-0.25);
\draw[lightgray] (3.25,0.25) -- (3.25,-0.25);
\draw[lightgray] (3.50,0.25) -- (3.50,-0.25);
\draw[lightgray] (3.75,0.25) -- (3.75,-0.25);
\draw[lightgray] (4.00,0.25) -- (4.00,-0.25);
\draw[lightgray] (4.25,0.25) -- (4.25,-0.25);
\draw[]          (4.50,0.25) -- (4.50,-0.25);
\draw[lightgray] (4.75,0.25) -- (4.75,-0.25);
\draw[lightgray] (5.00,0.25) -- (5.00,-0.25);
\draw[lightgray] (5.25,0.25) -- (5.25,-0.25);
\draw[]          (5.50,0.25) -- (5.50,-0.25);
\draw[lightgray] (5.75,0.25) -- (5.75,-0.25);
\draw[lightgray] (6.00,0.25) -- (6.00,-0.25);
\draw[lightgray] (6.25,0.25) -- (6.25,-0.25);
\draw[thick]     (6.50,0.25) -- (6.50,-0.25);
\draw[thick]     (7.00,0.25) -- (7.00,-0.25);
\draw[thick]     (7.25,0.25) -- (7.25,-0.25);
\draw[thick]     (7.75,0.25) -- (7.75,-0.25);
\draw[]          (8.00,0.25) -- (8.00,-0.25);
\draw[thick]     (9.33,0.25) -- (9.33,-0.25);
\draw[] (-0.25,0.25) -- (9.83,0.25);
\draw[] (-0.25,-0.25) -- (9.83,-0.25);
\draw (0.0,-0.025) node [] {$\cdots$};
\draw (0.375,-0.03) node [] {1};
\draw (1.5,-0.03) node [] {\contour{white}{probability}};
\draw (3.5,-0.03) node [] {\contour{white}{reward value}};
\draw (5.0,-0.03) node [] {\contour{white}{part.}};
\draw (6.0,-0.03) node [] {\contour{white}{index}};
\draw (6.775,-0.025) node [] {$\cdots$};
\draw (7.125,-0.03) node [] {2};
\draw (7.525,-0.025) node [] {$\cdots$};
\draw (7.875,-0.03) node [] {3};
\draw (8.67,-0.03) node [] {\contour{white}{is target?}};
\draw (9.63,-0.025) node [] {$\cdots$};
\draw[decorate] (0.25,0.3) -- (6.50,0.3);
\draw (3.375,0.65) node [] {branch ($\leq\,$25 bytes)};
\draw[decorate,decoration={mirror}] (6.95,-0.3) -- (7.3,-0.3);
\draw (7.125,-0.7) node [] {transition (1 byte)};
\draw[decorate] (7.75,0.3) -- (9.33,0.3);
\draw (8.54,0.65) node [] {state (2 bytes)};
\end{tikzpicture}
\caption{Inverse-sequential format to compactly represent MDP on disk}
\label{fig:SparseMatrixDisk}
\end{figure}
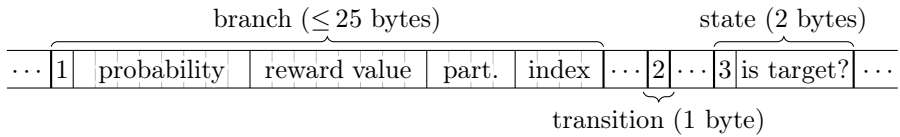

The format of \Cref{fig:SparseMatrixMemory} allows fast random access to all parts of the state space.
However, when only sequential access is required, an MDP can be stored more compactly.
\Cref{fig:SparseMatrixDisk} shows the ``inverse-sequential'' format used by our technique to store state spaces on disk.
States, transitions and branches are stored as a sequence of records, with the type of each record given by its first byte.
Branches can be stored even more compactly by adding record types for common cases such as branches with probability~1.
The key idea of the format is to first store all the branches of a transition before the transition record itself, and similarly store all the transitions (each preceded by its branches) of a state before the state record itself.
In this way, we do not need to store the number of transitions and the index of the first transition for a state since its transitions are precisely those that appeared since the previous state record (and analogously for the branches of a transition).
The random-access format of \Cref{fig:SparseMatrixMemory} can be reconstructed from a single sequential read of a file in the inverse-sequential format, and the file can be created sequentially with one simultaneous sequential pass through the arrays of the random-access format in memory.

\subsection{Disk-Based Exploration using Partitioning}

\begin{algorithm}[tp]
$\text{int }\mathit{count} := 1$, $\mathit{queue}_\mathrm{D}^1\text{.append}(s_0)$\;
\DoWhile{$\mathit{changed}$}{
  $\mathit{changed} := \False$\;
  \vspace{2mm}\tcp{iterate over all partitions discovered so far}
  \For{$i := 1$ to $\mathit{count}$\label{alg:PartitionedExploration:PartitionIteration}}{
    \vspace{2mm}\tcp{Phase 1: update preliminary target indices for cross transitions}
    \lForEach{$j \in \mathit{successors}^i$}{$\text{array }\mathit{updates}_\mathrm{M}^j := \mathit{updates}_\mathrm{D}^j\text{.load}()$\label{alg:PartitionedExploration:UpdateFirst}}
    $\mathit{oldmatrix}_\mathrm{D}^i := \mathit{matrix}_\mathrm{D}^i$, $\mathit{matrix}_\mathrm{D}^i\text{.clear}()$\tcp*{rename file}
    \ForEach(\tcp*[f]{read records sequentially}){$r \in \mathit{oldmatrix}_\mathrm{D}^i$}{
      \If{$r = \tuple{1, p, r, j, k} \wedge k < 0$}{$\mathit{matrix}_\mathrm{D}^i\text{.append}(\tuple{1, p, r, j, \mathit{updates}_\mathrm{M}^j[-k-1]})$\tcp*{update index}}
      \lElse{$\mathit{matrix}_\mathrm{D}^i\text{.append}(r)$}
    }
    $\text{unload }\mathit{updates}_\mathrm{M}^j\text{ for all }j \in \mathit{successors}^i$\label{alg:PartitionedExploration:UpdateLast}\;
    \vspace{3mm}\tcp{Phase 2: explore more states in breadth-first manner}
    $\mathit{updates}_\mathrm{D}^i\text{.clear}()$\label{alg:PartitionedExploration:UpdatesClear}\label{alg:PartitionedExploration:BFSFirst}\;
    $\text{queue }\mathit{queue}_\mathrm{M}^i := \mathit{queue}_\mathrm{D}^i\text{.load}()$, $\mathit{queue}_\mathrm{D}^i\text{.clear}()$, $\mathit{qlen}^i := 0$\;
    $\text{indexed-set }\mathit{states}_\mathrm{M}^i := \mathit{states}_\mathrm{D}^i\text{.load}()$\label{alg:PartitionedExploration:LoadStates}\;
    $\text{set }\mathit{done}^i := \mathit{states}_\mathrm{M}^i$\;
    \While{$\mathit{queue}_\mathrm{M}^i\text{.length} > 0$}{
      $\text{explicit-state }s := \mathit{queue}_\mathrm{M}^i\text{.dequeue}()$\;
      \lIf{$s \notin \mathit{states}_\mathrm{M}^i$}{$\mathit{states}_\mathrm{M}^i\text{.add}(s)$, $\mathit{states}_\mathrm{D}^i\text{.append}(s)$\label{alg:PartitionedExploration:AddState1}}
      $\mathit{updates}_\mathrm{D}^i\text{.append}(\mathit{states}_\mathrm{M}^i\text{.indexof}(s))$\label{alg:PartitionedExploration:UpdatesAppend}\;
      \leIf{$s \in \mathit{done}^i$}{$\Continue$}{$\mathit{changed} := \True$}
      \ForEach{$t \in s\text{.transitions}()$}{
        \lIf{$\neg\,t.guard(s.v)$}{\Continue}
        \ForEach{$b \in t\text{.branches}()$}{
          $\text{double }p := b\text{.probability}(s.v)$, $r := b\text{.reward}(s.v)$\;
          \lIf{$p = 0$}{\Continue}
          $\text{explicit-state }s' := b\text{.target}(s.v)$\;
          \If(\tcp*[f]{local transition}){$f(s') = i$}{
          \lIf{$s' \notin \mathit{states}_\mathrm{M}^i$}{$\mathit{states}_\mathrm{M}^i\text{.add}(s')$, $\mathit{states}_\mathrm{D}^i\text{.append}(s')$\label{alg:PartitionedExploration:AddState2}}
            $\mathit{queue}_\mathrm{M}^i\text{.enqueue}(s')$\label{alg:PartitionedExploration:LocalEnqueue}\;
            $\mathit{matrix}_\mathrm{D}^i\text{.append}(\tuple{1, p, r, i, \mathit{states}_\mathrm{M}^i\text{.indexof}(s')})$\label{alg:PartitionedExploration:AppendBranchLocal}\;
          }
          \Else(\tcp*[f]{cross transition}){
            $j := f(s')$, $\mathit{successors}^i\text{.add}(j)$, $\mathit{count} := \max\,\set{ \mathit{count}, j }$\;
            $\mathit{queue}_\mathrm{D}^j\text{.append}(s')$, $\mathit{qlen^j} = \mathit{qlen}^j + 1$\label{alg:PartitionedExploration:CrossEnqueue}\;
            $\mathit{matrix}_\mathrm{D}^j\text{.append}(\tuple{1, p, r, j, -\mathit{qlen}^j})$\label{alg:PartitionedExploration:AppendBranchCross}\tcp*{$\text{prelim.\ index} < 0$}
          }
        }
        $\mathit{matrix}_\mathrm{D}^i\text{.append}(\tuple{2})$\label{alg:PartitionedExploration:AppendTransition}\;
      }
      $\mathit{matrix}_\mathrm{D}^i\text{.append}(\tuple{3, s \in F})$\label{alg:PartitionedExploration:AppendState}\;
      $\mathit{done}^i\text{.add}(s)$\;
    }
    $\text{unload }\mathit{queue}_\mathrm{M}^i,\,\mathit{states}_\mathrm{M}^i,\,\mathit{done}^i$\label{alg:PartitionedExploration:UnloadSearchData}\label{alg:PartitionedExploration:BFSLast}\;
  }
}
\caption{\small Partitioned disk-based exploration with sparse matrix creation}
\label{alg:PartitionedExploration}
\end{algorithm}

Our disk-based exploration technique is given as \Cref{alg:PartitionedExploration}.
It is based on the approach of~\cite{BJ05,EK13}.
Files on disk are indicated by subscript~$_D$; when loaded into memory, the corresponding variable has subscript~$_M$.
For each partition, we use BFS to discover new states (lines~\ref{alg:PartitionedExploration:BFSFirst} to~\ref{alg:PartitionedExploration:BFSLast}) with the following data in memory:
\begin{itemize}
\item $\mathit{states}^i$:
The set of states (explicit-state representation) of partition~$i$ is loaded into memory in its entirety when search begins for the partition (line~\ref{alg:PartitionedExploration:LoadStates}).
States are added in memory and appended on disk (lines~\ref{alg:PartitionedExploration:AddState1} and~\ref{alg:PartitionedExploration:AddState2}).
\item $\mathit{queue}^i$:
The queue of states to explore in partition~$i$.
When a cross-transition is found during search in partition~$i$, \ie a branch leads to another partition $j \neq i$, then the target state is appended to $\mathit{queue}_\mathrm{D}^j$ on disk (line~\ref{alg:PartitionedExploration:CrossEnqueue}).
For local transitions, the target state is appended to $\mathit{queue}_\mathrm{M}^i$ in memory (line~\ref{alg:PartitionedExploration:LocalEnqueue}).
\item $\mathit{done}^i$:
The in-memory set of fully explored states for the current iteration.
\end{itemize}
When an iteration of search in partition~$i$ ends, $\mathit{states}^i$ is backed on disk, $\mathit{queue}^i$ is empty, and $\mathit{done}^i$ is no longer needed, so we remove them from memory (line~\ref{alg:PartitionedExploration:UnloadSearchData}).

During search, we simultaneously create the sparse matrix-like representation of the partitions on disk in files $\mathit{matrix}_\mathrm{D}^i$ using the inverse-sequential format.
The files are not loaded into memory.
The records for new branches, transitions and states are appended to the file in lines~\ref{alg:PartitionedExploration:AppendBranchLocal}, \ref{alg:PartitionedExploration:AppendBranchCross}, \ref{alg:PartitionedExploration:AppendTransition} and~\ref{alg:PartitionedExploration:AppendState}.
The main complication is the correct treatment of cross transitions:
A branch record stores the partition~$j$ of its target state~$s'$ and the index of~$s'$ within that partition.
However, we cannot determine this index without loading all of $\mathit{states}_\mathrm{D}^j$ into memory, and even then, $s'$~may not have been explored yet.
To solve this problem, we instead use the index of $s'$ in $\mathit{queue}_\mathrm{D}^j$, which is easily determined (line~\ref{alg:PartitionedExploration:CrossEnqueue}).
To distinguish such a preliminary index, which needs to be corrected later, from a local or already corrected one, we store it as a negative value (line~\ref{alg:PartitionedExploration:AppendBranchCross}).

The correction of these preliminary indices inside $\mathit{matrix}_\mathrm{D}^i$ happens at the beginning of an iteration for partition~$i$ (lines~\ref{alg:PartitionedExploration:UpdateFirst} to~\ref{alg:PartitionedExploration:UpdateLast}).
The files $\mathit{updates}_\mathrm{D}^j$ for all successor partitions~$j$ are loaded into memory.
These files have been created by the previous iteration for partition~$j$ in lines~\ref{alg:PartitionedExploration:UpdatesClear} and~\ref{alg:PartitionedExploration:UpdatesAppend} and contain the correct indices for all states that were previously in $\mathit{queue}_\mathrm{D}^j$, at the same position.
The preliminary queue-based indices in partition~$i$ can thus be corrected by a sequential pass through its sparse matrix-like representation in file $\mathit{matrix}_\mathrm{D}^i$, replacing all negative indices $-k$ for partition~$j$ by the corrected value at $\mathit{updates}_\mathrm{M}^j[k]$.
This is a random-access operation on the files $\mathit{updates}_\mathrm{D}^j$, which is why they were loaded into memory beforehand, but a sequential operation on the file $\mathit{matrix}_\mathrm{D}^i$, of which we thus only need to load into memory one record at a time.
Observe that this correction process relies on the availability of $\mathit{updates}_\mathrm{D}^j$ for all successor partitions~$j$.
To assure this, we iterate over all partitions in a fixed order in line~\ref{alg:PartitionedExploration:PartitionIteration} instead of always moving to the partition with the longest queue as in~\cite{BJ05,EK13}.

To describe the memory usage and I/O complexity of this algorithm, let $n_{\max}$ denote the max.\ number of states, $s_{\max}$ the max.\ number of successor partitions (\ie the max.\ outdegree of the partition graph), and $c_{\max}$ the max.\ number of \emph{incoming} cross edges, over all partitions.
Then the correction of preliminary indices in phase~1 needs memory in $O(s_{\max} \cdot c_{\max})$ for the $\mathit{updates}_\mathrm{M}^j$ arrays and the exploration in phase~2 needs memory in $O(n_{\max} + c_{\max})$ for $\mathit{states}_\mathrm{M}^i$ and $\mathit{done}^i$ plus $\mathit{queue}_\mathrm{M}^i$.
Additionally, we need memory for the sets of integers $\mathit{successors}^i$, which we assume to be negligible compared to the other data items.
A theoretical analysis of the I/O complexity~\cite{AV88} of a partitioning-based technique is problematic (and in fact absent from~\cite{BJ05} and~\cite{EK13}) due to the way multiple files are used \eg when cross transitions are encountered:
For the (unusual)  case of very small $n_{\max}$ and very high $s_{\max}$ and $c_{\max}$, the disk accesses to append target states to different queues would be mostly random, but in practice (with low $s_{\max}$ and I/O buffering) they are almost purely sequential.
A theoretical worst-case analysis would thus be too pessimistic to be useful.
We consequently abstain from such an analysis, too, and rely on the experimental evaluation of \Cref{sec:Evaluation}.

However, it is clear that the structure of the model \wrt the partitioning function will have a high impact on performance in general; in particular, a low number of cross edges is most desirable for the exploration algorithm presented here.
Ideally, the partition graph is also forward-acyclic.
In that case, two iterations of the outermost loop suffice:
All states are explored in the first iteration, and the second only corrects the preliminary indices.

\section{Disk-Based Partitioned Value Iteration}
\label{sec:PartitionedVI}

\begin{algorithm}[t]
\For(\tcp*[f]{prepare value arrays on disk}){$i := 1$ to $\mathit{count}$\label{alg:ValueIterationPartitioned:PrepareStart}}{
  $\mathit{matrix}_\mathrm{M}^i := \mathit{matrix}_\mathrm{D}^i\text{.load}()$\;
  \For{$k := 0$ to $\mathit{matrix}_\mathrm{M}^i\text{.states.length} - 1$}{
    $\mathit{values}_\mathrm{D}^i\text{.append}(${$\mathit{matrix}_\mathrm{M}^i\text{.states}[k]\text{.istarget}$} ? {$1$} : {$0$}$)$\;
  }
  $\text{unload }\mathit{matrix}_\mathrm{M}^i$\label{alg:ValueIterationPartitioned:PrepareEnd}\;
}
\While(\tcp*[f]{block-iterative value iteration}\label{alg:ValueIterationPartitioned:IterateStart}){$\mathit{changed}$}{
  $\mathit{changed} := \False$\tcp*{changed is initially false}
  \For{$i := count$ down to $\mathit{1}$\label{alg:ValueIterationPartitioned:Iterate}}{
    $\mathit{matrix}_\mathrm{M}^i := \mathit{matrix}_\mathrm{D}^i\text{.load}()$, $\mathit{values}_\mathrm{M}^i := \mathit{values}_\mathrm{D}^i\text{.load}()$\label{alg:ValueIterationPartitioned:LoadLocal}\;
    \lForEach{$j \in \mathit{successors^i}$}{$\mathit{values}_\mathrm{M}^j := \mathit{values}_\mathrm{D}^j\text{.load}()$\label{alg:ValueIterationPartitioned:LoadCross}}
    \Repeat{$\mathit{error} < \epsilon$}{
      $\mathit{error} := 0$\;
      \For{$k := 0\text{ to }\mathit{matrix}_\mathrm{M}^i\text{.states.length} - 1$}{
        \lIf{$\mathit{matrix}_\mathrm{M}^i\text{.states}[k]\text{.istarget}$}{\Continue}
        $v_\mathit{new} := \max \ldots$\label{alg:ValueIterationPartitioned:Update}\tcp*{as in \Cref{alg:ValueIteration}, but with $\mathit{values}_\mathrm{M}^i$/$\mathit{values}_\mathrm{M}^j$}
        \lIf(\tcp*[f]{compute error as in \Cref{alg:ValueIteration}}){$v_\mathit{new} > 0$}{$\mathit{error} := \ldots$}
        $\mathit{values}_\mathrm{M}^i[k] := v_\mathit{new}$\;
        \lIf{$\mathit{error} \geq \epsilon$}{$\mathit{changed} := \True$}
      }
    }
    $\text{unload }\mathit{matrix}_\mathrm{M}^i,\,\mathit{values}_\mathrm{M}^i\text{ and the }\mathit{values}_\mathrm{M}^j\text{ for all }j \in \mathit{successors}^i$\label{alg:ValueIterationPartitioned:IterateEnd}\;
  }
}
\Return{$\mathit{values}_\mathrm{D}^1[0]$}
\caption{\small Partitioned value iteration for max.\ reachability probabilities}
\label{alg:ValueIterationPartitioned}
\end{algorithm}

The result of the partitioned exploration presented in the previous section is a set of files in inverse-sequential format for the partitions of the state space.
As mentioned in \Cref{sec:Introduction}, value iteration can update the states in any order, as long as the maximum error for termination is computed in a way that takes all states into account.
We can thus apply value iteration in a block-iterative manner to the partitions of the state space as shown in \Cref{alg:ValueIterationPartitioned}.
The vector of values for each partition is stored in a separate file on disk.
In lines~\ref{alg:ValueIterationPartitioned:PrepareStart} to~\ref{alg:ValueIterationPartitioned:PrepareEnd}, these files are created with the initial values based on whether a state is in the target set~$F$.
The actual value iterations are then performed in lines~\ref{alg:ValueIterationPartitioned:IterateStart} to~\ref{alg:ValueIterationPartitioned:IterateEnd}.
For each partition, we need to load the sparse matrix-style representation of this part of the MDP into memory in the random-access format of \Cref{fig:SparseMatrixMemory}, plus the values for the current partition (line~\ref{alg:ValueIterationPartitioned:LoadLocal}), and those of its successors (lines~\ref{alg:ValueIterationPartitioned:LoadCross}).
The values of the successor partitions are needed to calculate the current state's new value in line~\ref{alg:ValueIterationPartitioned:Update} in presence of cross transitions.
Memory usage is thus in $O(m_{\max} + s_{\max} \cdot n_{\max})$, where $m_{\max}$ is the maximum over all partitions of the sum of the number of states, transitions and branches.
The I/O complexity is in $O(i \cdot p \cdot (\mathrm{scan}(m_{\max}) + (s_{\max} + 1) \cdot \mathrm{scan}(n_{\max})))$ where $i$ is the number of iterations of the outermost loop starting in line~\ref{alg:ValueIterationPartitioned:IterateStart} and $p$ is the total number of partitions.

In contrast to the exploration phase, the performance of this disk-based value iteration is not directly affected by the number of cross transitions.
However, the number of successor partitions, \ie $s_{\max}$, is crucial.
An additional consideration is the way that values propagate through the partitions.
The ideal case is again a forward-acyclic partition graph, for which
a single iteration of the outermost loop (line~\ref{alg:ValueIterationPartitioned:IterateStart}) suffices since we iterate over the partitions in reverse order (line~\ref{alg:ValueIterationPartitioned:Iterate}).

For expected rewards, we additionally need to precompute the sets of states that reach the target set with probability one and zero as mentioned in \Cref{sec:Preliminaries}.
The standard graph-based fixpoint algorithms used for this purpose~\cite{FKNP11} can be changed to work in a block-iterative manner in the same way as value iteration.

\section{Evaluation}
\label{sec:Evaluation}

In this section, we investigate the behaviour of our disk-based probabilistic model checking approach and its implementation in \mcsta on five models from the literature. 
%
Experiments were performed on an Intel Core i7-4650U system with 8\sunit{GB} of memory and a 2\sunit{TB} USB~3.0 magnetic hard disk, running
64-bit Windows 8.1 for \mcsta and Ubuntu Linux 14.10 for \prism version 4.2.1.
We used a timeout of 12~hours.
Memory measurements refer to peak working/resident sets.
Since \mcsta (implemented in C\#) and parts of \prism are garbage-collected, however, the reported memory usages may fluctuate and be higher than what is actually necessary to solve the task at hand.
Our experiments show what the disk-based approach makes possible on standard workstation configurations today; by using compute servers with more memory, we can naturally scale to even larger models.

Detailed performance results are shown in \Cref{tab:Evaluation}.
State space sizes are listed in \emph{millions} of states, so the largest model has about 3.5~\emph{billion} states.
Columns ``exp'' and ``chk'' show the runtime of the exploration and analysis phases, respectively, in \emph{minutes}.
Columns ``\ssunit{\!GB}'' list the peak memory usage over both phases in \emph{gigabytes}.
We show the performance of \mcsta without using the disk to judge the overhead of partitioning and disk usage. 
Where possible, we also compare with \prism, which does not use the disk, but provides a semi-symbolic \textsc{hybrid} engine that uses BDD to compactly represent the states, transitions and branches while keeping the entire value vector(s) in memory during value iteration (limiting its scalability), and a fully symbolic \textsc{mtbdd} engine that also uses BDD for the value vector.
The \textsc{hybrid} engine does not support expected rewards.

\begin{table}[p]
\caption{Evaluation results (millions of states, minutes, and gigabytes of memory)}
\label{tab:Evaluation}
\setlength{\tabcolsep}{2.5pt}
\centering%
\makebox[\textwidth][c]{\begin{tabular}{ccr|rrr|rr;{0.4pt/1.6pt}rrr|rrr|rrr}
\multicolumn{3}{c|}{model} & \multicolumn{3}{c|}{\parbox{1.6cm}{\centering in-memory\\[-1.5mm]\begin{scriptsize}(\mcsta)\end{scriptsize}}} & \multicolumn{5}{c|}{\parbox{1.8cm}{\centering \textbf{disk-based\\[-1.25mm]\begin{scriptsize}(\mcsta \textsc{-l})\end{scriptsize}}}} & \multicolumn{3}{c|}{\parbox{2.1cm}{\centering semi-symbolic\\[-1.25mm]\begin{scriptsize}(\prism \textsc{hybrid})\end{scriptsize}}} & \multicolumn{3}{c}{\parbox{2.1cm}{\centering fully symbolic\\[-1.25mm]\begin{scriptsize}(\prism \textsc{mtbdd})\end{scriptsize}}} \\
&params & states & exp & chk & \ssunit{GB} & $p\,$ & $n_{\max}\!\!$ & exp & chk & \ssunit{GB} & exp & chk & \ssunit{GB} & exp & chk & \ssunit{GB} \\\hline

\multirow{6}{*}{\rotatebox[origin=c]{90}{$\text{CSMA/CD}_{1\times\text{P}}^{N\!,K}$}}
 &3,\,4 & 1.5 & 0.1 & 0.0 & {0.3} & 12 & 0.4 & 0.2 & 0.0 & 0.2 & 0.0 & 0.2 & 0.2 & 0.0 & 0.5 & 0.3 \\
&3,\,5 & 12.1 & 1.1 & 0.1 & 2.6 & 15 & 2.6 & 1.3 & 0.1 & 0.7 & 0.1 & 1.6 & 0.5 & 0.1 & 4.0 & 0.4 \\
&3,\,6 & 84.9 & \multicolumn{3}{c|}{$> 8\sunit{GB}$} & 18 & 15.3 & 9.3 & 1.3 & 5.0 & 0.3 & 13.1 & 2.3 & 0.3 & 22.9 & 3.0 \\\cdashline{2-17}[0.4pt/1.6pt]
&4,\,3 & 8.2 & 1.0 & 0.1 & 1.7 & 12 & 2.7 & 1.1 & 0.1 & 0.8 & 0.1 & 0.8 & 0.4 & 0.1 & 2.2 & 0.4 \\
&4,\,4 & 133.3 & \multicolumn{3}{c|}{\multirow{2}{*}{$> 8\sunit{GB}$}} & 16 & 33.0 & 19.1 & 2.2 & 6.6 & 0.4 & 17.6 & 3.6 & 0.6 & 21.7 & 5.1 \\
&4,\,5 & 2596.0 & && & & & \multicolumn{3}{c|}{$> 8\sunit{GB}$} & \multicolumn{3}{c|}{$> 8\sunit{GB}$} & \multicolumn{3}{c}{$> 12\sunit{h}$} \\\hline

\multirow{6}{*}{\rotatebox[origin=c]{90}{$\text{CSMA/CD}_{2\times\text{E}}^{N\!,K}$}}
 &3,\,4 & 1.5 & 0.1 & 0.1 & 0.3 & 12 & 0.4 & 0.2 & 0.2 & 0.2 & \multicolumn{3}{c|}{\multirow{3}{*}{n/a}} & 0.0 & 18.1 & 0.4 \\
&3,\,5 & 12.1 & 1.1 & 1.5 & 2.6 & 15 & 2.6 & 1.3 & 1.7 & 0.7 & && & 0.1 & 96.9 & 4.7 \\
&3,\,6 & 84.9 & \multicolumn{3}{c|}{$> 8\sunit{GB}$} & 18 & 15.3 & 9.3 & 19.4 & 5.0 & && & 0.3 & 707.0 & 5.1 \\\cdashline{2-17}[0.4pt/1.6pt]
&4,\,3 & 8.2 & 1.0 & 0.9 & 1.7 & 12 & 2.7 & 1.1 & 0.9 & 0.8 & \multicolumn{3}{c|}{\multirow{3}{*}{n/a}} & 0.1 & 92.4 & 0.5 \\
&4,\,4 & 133.3 & \multicolumn{3}{c|}{\multirow{2}{*}{$> 8\sunit{GB}$}} & 16 & 33.0 & 19.1 & 16.5 & 6.6 & && & 0.5 & 637.3 & 5.5 \\
&4,\,5 & 2596.0 & && & & & \multicolumn{3}{c|}{$> 8\sunit{GB}$} & && & \multicolumn{3}{c}{$>12\sunit{h}$} \\\hline

\multirow{6}{*}{\rotatebox[origin=c]{90}{$\text{Consensus}_{2\times\text{P}}^{N\!,K}$}}
 &8,\,2 & 61.0 & \multicolumn{3}{c|}{\multirow{6}{*}{$> 8\sunit{GB}$}} & 5 & 16.8 & 10.5 & 104.9 & 6.4 & 0.0 & 28.3 & 1.6 & 0.0 & 5.4 & 0.3 \\
&8,\,3 & 87.9 & & & & 7 & 16.8 & 16.0 & 200.6 & 4.3 & 0.0 & 65.1 & 2.3 & 0.0 & 10.1 & 0.4 \\
&8,\,4 & 114.8 & & & & 8 & 16.8 & 21.8 & 347.5 & 7.3 & 0.0 & 121.4 & 2.9 & 0.0 & 17.5 & 0.4 \\
&8,\,5 & 141.6 & & & & 10 & 16.8 & 27.2 & 484.9 & 6.8 & 0.0 & 193.4 & 3.6 & 0.0 & 25.1 & 0.4 \\
&8,\,6 & 168.5 & & & & 12 & 16.8 & 33.9 & 660.3 & 6.9 & 0.0 & 260.6 & 4.2 & 0.0 & 38.9 & 0.4 \\
&8,\,7 & 195.4 & & & & & & \multicolumn{3}{c|}{$> 12\sunit{h}$} & 0.0 & 361.6 & 4.9 & 0.0 & 49.9 & 0.4 \\\hline

\multirow{4}{*}{\rotatebox[origin=c]{90}{$\text{WLAN}_{1\!\times\!\text{P}}^{K}$}}
 &1 & 718.0 & \multicolumn{3}{c|}{\multirow{4}{*}{$> 8\sunit{GB}$}} & 203 & 11.5 & 177.3 & 8.5 & 3.0 & \multicolumn{3}{c|}{\multirow{4}{*}{$> 8\sunit{GB}$}} & 715.3 & 4.3 & 5.8 \\
&2 & 1197.9 & & & & 337 & 12.0 & 283.5 & 15.7 & 3.0 & & & & \multicolumn{3}{c}{\multirow{3}{*}{$> 12\sunit{h}$}} \\
&3 & 1685.0 & & & & 471 & 13.1 & 392.2 & 23.4 & 3.0 & & & & & & \\
&4 & 2186.7 & & & & 605 & 15.1 & 502.6 & 30.7 & 3.5 & & & & & & \\\hline

\multirow{4}{*}{\rotatebox[origin=c]{90}{$\text{WLAN}_{1\!\times\!\text{E}}^{K}$}}
 &1 & 718.0 & \multicolumn{3}{c|}{\multirow{4}{*}{$> 8\sunit{GB}$}} & 203 & 11.5 & 177.3 & 52.4 & 3.0 & \multicolumn{3}{c|}{\multirow{4}{*}{n/a}} & \multicolumn{3}{c}{\multirow{4}{*}{$> 12\sunit{h}$}} \\
&2 & 1197.9 & & & & 337 & 12.0 & 283.5 & 72.0 & 3.0 & & & & & & \\
&3 & 1685.0 & & & & 471 & 13.1 & 392.2 & 94.2 & 3.0 & & & & & & \\
&4 & 2186.7 & & & & 605 & 15.1 & 502.6 & 114.0 & 3.5 & & & & & & \\\hline

\multirow{4}{*}{\rotatebox[origin=c]{90}{$\text{BRP}_{6\times\text{P}}^{N\!,\mathit{T\!D}}$}}
 &\phantom{1}64,\,16 & 18.7 & 1.5 & 0.2 & 3.8 & 65 & 0.3 & 1.8 & 0.5 & 0.2 & 23.0 & 56.8 & 1.0 & \multicolumn{3}{c}{\multirow{2}{*}{error}} \\
&128,\,16 & 37.4 & 3.1 & 0.5 & 7.3 & 129 & 0.3 & 3.7 & 0.9 & 0.2 & 34.7 & 150.4 & 1.4 & & & \\\cdashline{2-17}[0.4pt/1.6pt]
&\phantom{1}64,\,32 & 70.7 & \multicolumn{3}{c|}{\multirow{2}{*}{$> 8\sunit{GB}$}} & 65 & 1.2 & 7.4 & 1.8 & 0.5 & 89.4 & 345.2 & 2.4 & \multicolumn{3}{c}{\multirow{2}{*}{error}} \\
&128,\,32 & 141.5 & & & & 129 & 1.2 & 15.3 & 3.4 & 0.5 & \multicolumn{3}{c|}{$> 12\sunit{h}$} & & & \\\hline

\multirow{4}{*}{\rotatebox[origin=c]{90}{$\text{BRP}_{2\times\text{TP}}^{N\!,D}$}}
 &\phantom{1}64,\,256 & 355.7 & \multicolumn{3}{c|}{\multirow{2}{*}{$> 8\sunit{GB}$}} & 577 & 1.5 & 40.7 & 3.3 & 0.6 & \multicolumn{3}{c|}{\multirow{2}{*}{$> 8\sunit{GB}$}} & 122.6 & 38.2 & 2.6 \\
&128,\,256 & 715.6 & & & & 1153 & 1.5 & 93.0 & 60.6 & 0.6 & & & & \multicolumn{3}{c}{$>12\sunit{h}$} \\\cdashline{2-17}[0.4pt/1.6pt]
&\phantom{1}64,\,512 & 1773.7 & \multicolumn{3}{c|}{\multirow{2}{*}{$> 8\sunit{GB}$}} & 1089 & 4.8 & 203.1 & 18.8 & 1.6 & \multicolumn{3}{c|}{\multirow{2}{*}{$> 8\sunit{GB}$}} & \multicolumn{3}{c}{\multirow{2}{*}{$> 8\sunit{GB}$}} \\
&128,\,512 & 3573.3 & & & & 2177 & 4.8 & 418.5 & 38.1 & 1.8 & & & & & & \\\hline

\multirow{8}{*}{\rotatebox[origin=c]{90}{$\text{File server}_{2\times\text{TP}}^{C\!,D}$}}
 &\phantom{1}5,\,100 & 18.0 & 1.4 & 0.5 & 5.4 & 102 & 0.2 & 2.0 & 0.4 & 0.2 & \multicolumn{3}{c|}{\multirow{4}{*}{n/a}} & \multicolumn{3}{c}{\multirow{4}{*}{n/a}} \\
&\phantom{1}5,\,200 & 41.2 & \multicolumn{3}{c|}{\multirow{3}{*}{$> 8\sunit{GB}$}} & 202 & 0.2 & 4.7 & 1.0 & 0.2 & & & & & & \\
&\phantom{1}5,\,400 & 87.8 & & & & 402 & 0.2 & 10.5 & 2.1 & 0.2 & & & & & & \\
&\phantom{1}5,\,800 & 180.9 & & & & 802 & 0.2 & 22.4 & 4.3 & 0.2 & & & & & & \\\cdashline{2-17}[0.4pt/1.6pt]
&10,\,100 & 34.0 & \multicolumn{3}{c|}{\multirow{4}{*}{$> 8\sunit{GB}$}} & 102 & 0.4 & 4.0 & 0.9 & 0.2 & \multicolumn{3}{c|}{\multirow{4}{*}{n/a}} & \multicolumn{3}{c}{\multirow{4}{*}{n/a}} \\
&10,\,200 & 77.1 & & & & 202 & 0.4 & 9.6 & 1.9 & 0.2 & & & & & & \\
&10,\,400 & 163.4 & & & & 402 & 0.4 & 20.4 & 4.1 & 0.3 & & & & & & \\
&10,\,800 & 335.9 & & & & 802 & 0.4 & 43.9 & 8.6 & 0.3 & & & & & & \\\hline

&params & states & exp & chk & \ssunit{GB} & $p\,$ & $n_{\max}\!\!$ & exp & chk & \ssunit{GB} & exp & chk & \ssunit{GB} & exp & chk & \ssunit{GB}
\end{tabular}}
\end{table}

\paragraph{Compression.}
As all file accesses are sequential, we can use generic lossless compression to reduce disk accesses.
Using the LZ4 algorithm~\cite{Web:LZ4}, we achieved a $7\times$ to $10\times$ reduction in disk usage on our examples.
We observed almost no change in runtime with compression enabled, so the extra CPU time is outweighed by reduced disk I/O.
Compression thus lowers disk usage at no runtime costs.

\paragraph{Partitioning functions.}
The actual performance of our approach depends on the structure of the model and its interplay with the partitioning function.
Scalability hinges on the function's ability to distribute the states such that the largest partition and the values of its successors fit into memory.
The problem of automatically constructing a good partitioning function has largely been solved in prior work, and many techniques, like the ones described and referenced in~\cite{EK13}, are available, but they are not yet implemented in \mcsta.
For our evaluation, we thus use relatively simple manually specified partitioning functions.

\subsubsection{CSMA/CD:}

The MDP model of the IEEE 802.3 CSMA/CD protocol from the \prism benchmark suite.
It was manually constructed from a PTA model using the digital clocks approach.
It has parameters $N$, the number of communicating nodes, and $K$, the maximum value of the backoff counter.
The nodes count the number of collisions they encounter when trying to send a message.
We partition according to the sum of the collision counters of the nodes.
The resulting partition graph is forward-acyclic since these counters are only incremented, and $s_{\max} = N$.
However, due to using the sum of several values for partitioning, the states are not evenly distributed over the partitions.

We first report on the performance of computing the minimum probability of any node eventually delivering its message with fewer than $K$ collisions (model $\text{CSMA/CD}_{1\times\text{P}}^{N\!,K}$ in \Cref{tab:Evaluation}, with $1\times\text{P}$ indicating that one reachability probability is computed), and then on computing the max.\ and min.\ expected times until all nodes have delivered their message (model $\text{CSMA/CD}_{2\times\text{E}}^{N\!,K}$, where $2\times\text{E}$ indicates that we compute two expected-reward values).
All MDP are only medium-sized.
Our disk-based technique achieves performance comparable to the semi-symbolic approach here, which however does not support expected rewards.
The fully symbolic approach has significantly higher runtimes for those properties.

\subsubsection{Randomised Consensus:}

The \prism benchmark of the randomised consensus protocol of $N$~actors doing random walks bounded by~$K$ to reach a common decision.
We partition according to the value of the shared counter variable.
The resulting partition graph is strongly connected with $s_{\max} = 2$.
We use $\epsilon = 0.02$ during value iteration (instead of the default $\epsilon = 10^{-6}$ as in the other examples).
The MDP appear medium-sized in terms of states, but have about $5\times$ as many transitions and $7\times$ as many branches as states, so should be considered large.

We check the two probabilistic reachability properties originally named ``$\text{C}_1$'' and ``$\text{C}_2$''.
The fully symbolic technique completes exploration and analysis much faster than our disk-based approach.
This is because this model is a benchmark for value iteration, with values propagating in very small increments back-and-forth through all the states and thus partitions.
Still, we observe that $n_{\max}$ is invariant under~$K$, so our technique will be able to check this model for $N = 8$ and any value of $K$ without running out of memory---if given enough time.

\subsubsection{Wireless LAN:}

The \modest PTA model~\cite{HH09} of IEEE 802.11 WLAN, based on~\cite{KNS02}.
So far, this protocol has only been analysed with reduced timing parameters to contain state space explosion.
We use the original values of the standard for a 2\,Mbps transmission rate instead, including the max.\ transmission time of $15717\sunit{$\upmu\text{s}$}$, with $1\sunit{$\upmu\text{s}$}$ as one model time unit.
Parameter $K$ is the maximum value of the backoff counter.
We partition according to the first station's backoff counter, its control location, and its clock.
The resulting partition graph has some cycles with $s_{\max} = 3$.
Exploration needs 5~iterations of the outermost loop of \Cref{alg:PartitionedExploration} in all cases.
We compute the maximum\ probability that either station's backoff counter reaches $K$ (model $\text{WLAN}_{1\times\text{P}}^{K}$ in \Cref{tab:Evaluation}) as well as the maximum\ expected time until one station delivers its packet ($\text{WLAN}_{1\times\text{E}}^{K}$).

\subsubsection{BRP:}

The \modest PTA model of the Bounded Retransmission Protocol (BRP) from~\cite{HH09}.
Parameters are $N$, the number of data frames to be transmitted, $\mathit{MAX}$, the bound on the retries per frame, and $\mathit{TD}$, the maximum transmission delay.
We fix $\mathit{MAX} = 12$.
We partition by the number of the current data frame to analyse the model's six probabilistic reachability properties ($\text{BRP}_{6\times\text{P}}^{N\!,\mathit{T\!D}}$).
This leads to the ideal case of a forward-acyclic partition graph with $s_{\max} = 1$.
We also analyse two time-bounded reachability properties ($\text{BRP}_{2\times\text{TP}}^{N\!,D}$) with deadline~$D$ and fixed $\mathit{TD} = 32$, partitioning additionally according to the values of the added global clock.
This leads to $s_{\max}=2$.
For the reachability probabilities, \prism's \textsc{mtbdd} engine incorrectly reported probability zero in all cases.
Our approach benefits hugely from having to perform far fewer total value iterations per state due to the favourable partitioning.
In the reachability probabilities case, $n_{\max}$ is invariant under $N$, so we can scale $N$ arbitrarily without running out of memory.

\subsubsection{File Server:}

The STA file server model from~\cite{HHH14}.
$C$ is the capacity of the request buffer.
We compute the maximum\ and the minimum\ probability of a buffer overflow within time bound~$D$.
We cannot compare with \prism because some features necessary to support STA cannot currently be translated into its input language from \modest. 
Using our disk-based technique permits a finer abstraction for continuous probability distributions than before ($\rho = 0.01$ instead of $0.05$).
We partition according to the values of the global clock introduced to check the time bounds.
This leads to the ideal case of an acyclic partition graph with $s_{\max} = 1$.
The state space and number of partitions grow linearly in the time bound while
$n_{\max}$ remains invariant.
We can thus check time-bounded properties for any large bound without exceeding the available memory, at a linear increase in runtime.
This solves a major problem in STA model checking.

\section{Conclusion}
\label{sec:Conclusion}

We have shown that the state space partitioning approach to using secondary storage for model checking combines well with analysis techniques built on graph fixpoint algorithms.
We have used the example of MDP models and value iteration, but the same scheme is applicable to other techniques, too.
In particular, the precomputation step for expected-reward properties is very close to what is needed for CTL model checking.
Our technique is implemented in the \mcsta tool of the \toolset, available at \href{http://www.modestchecker.net/}{\texttt{www.modestchecker.net}}.
In our evaluation, we observed that it significantly extends the reach of probabilistic model checking.
It appears complementary to the symbolic approach:
On the model where our technique struggles, \prism performs well, and where \prism runs into memory or time limitations, our technique appears to work well.
In particular, our approach appears to work better for expected-reward properties, and we have been able to defuse the crippling state space explosion caused by the deadlines of time-bounded reachability properties in PTA and STA models.

\bibliography{paper}{}
\bibliographystyle{splncs03}

\end{document}